\documentclass[a4paper,11pt]{article}
\pdfoutput=1 

\usepackage{jcappub} 

\usepackage[T1]{fontenc} 
\usepackage{feynmp}
\usepackage{gensymb}

\title{JUNO Sensitivity to Resonant Absorption of Galactic Supernova Neutrinos by Dark Matter}

\author{Tarso Franarin,}
\author{Malcolm Fairbairn}
\author{and Jonathan H. Davis}

\affiliation{Theoretical Particle Physics and Cosmology, Department of Physics,
King's College London, London WC2R 2LS, United Kingdom}

\emailAdd{tarso.franarin@kcl.ac.uk}
\emailAdd{jonathan.davis@kcl.ac.uk}
\emailAdd{malcolm.fairbairn@kcl.ac.uk}

\abstract{Resonant interactions between neutrinos from a Galactic supernova and dark matter particles can lead to a sharp dip in the neutrino energy spectrum. Due to its excellent energy resolution, measurement of this effect with the JUNO experiment can provide evidence for such couplings. We discuss how JUNO may confirm or further constrain a model where scalar dark matter couples to active neutrinos and another fermion.  }

\begin{document}
\maketitle
\flushbottom

\section{Introduction}

The nature of dark matter (DM) and the origin of neutrino masses both point to physics beyond the Standard Model (SM). Neutrinos are known to have interactions other than gravitational, while many models contain a coupling between DM and SM particles in order to explain the production mechanism in the early Universe. Both particles are elusive, dark matter has so far completely evaded detection while neutrinos are so weakly interacting they are very difficult to study. Because of this we are somewhat free to postulate the nature of their interactions and in this theme there are many works in the literature coupling dark matter with neutrinos, \cite{Krauss:2002px,Cheung:2004xm,Asaka:2005an,Ma:2006km,Kubo:2006yx,Ma:2006fn,Hambye:2006zn,Chun:2006ss,Kubo:2006rm,Boehm:2006mi,Farzan:2009ji,Farzan:2010mr,Farzan:2010wh,Farzan:2011tz,Davoudiasl:2018hjw} being just a relevant subset of a huge number of works.

Core-collapse supernova (SN) explosions, being abundant sources of neutrinos which have to pass through our galaxy before arriving on Earth, can provide information about DM-neutrino interactions. We expect at least a few Galactic supernovae per century \cite{Ando:2005ka}, and large neutrino detectors should obtain high-statistics data of the supernova neutrino spectrum when the next explosion occurs \cite{Scholberg:2012id,Lu:2016ipr,Horiuchi:2017sku,Nikrant:2017nya,Li:2017dbg}.

It was pointed out nearly 40 years ago that absorption of ultra-high-energy neutrinos ($E_\nu > 10^{21}$ eV) on cosmic relic neutrinos would lead to a dip on the spectrum at energies corresponding to formation of Z bosons \cite{Weiler:1982qy,Roulet:1992pz,Yoshida:1996ie,Eberle:2004ua,Barenboim:2004di}. Similarly, in models with additional light $Z^\prime$ gauge bosons coupled to neutrinos, the same feature could be obtained for lower energy neutrinos such as those produced in supernovae \cite{Goldberg:2005yw,Baker:2006gm}. The flux of high-energy neutrinos could also be distorted by interactions with ultralight scalar dark matter \cite{Reynoso:2016hjr}. 

Our framework is inspired by the work of Farzan and Palomares-Ruiz in which the coupling of neutrinos with dark matter gives rise to a dip in the diffuse supernova neutrino background \cite{Farzan:2014gza}. When propagating cosmological distances, the redshift of neutrinos becomes important and leads to a broadened dip. Here we consider the same model and analyse the effects of such coupling on the flux of neutrinos from a Galactic supernova. Given the narrowness of the absorption feature, we consider the future JUNO detector, designed to have excellent energy resolution. 

The outline of this paper is as follows. In Section \ref{sec:model} we describe the model and different constraints on its parameters. In section \ref{sec:distortion} we discuss the distortion of the energy spectrum of supernova neutrinos due to their interactions with dark matter particles. In Section \ref{sec:detection} we consider the detection of these neutrinos by JUNO. We examine our results in Section \ref{sec:results} and summarise our findings in Section \ref{sec:conclusions}.

\section{Dark matter interactions with neutrinos}
\label{sec:model}

In this model the neutrino vacuum mass eigenstates, $\nu_i$ ($i$ = 1, 2, 3), couple to a new scalar $\phi$ and a new fermion $N$ via
\begin{equation}
g_iN_R^\dagger\nu_{i,L}\phi,
\label{eq:coupling}
\end{equation}
where $g_i=\sum_\alpha U_{\alpha i} g_\alpha$. The lightest of the particles $\phi$ and $N$ is a DM candidate if the complete Lagrangian is invariant under a $Z_2$ symmetry. Depending on the proprieties of $\phi$ and $N$ we can identify different scenarios: $\phi$ can be real or complex, $N$ can be of Dirac or Majorana type and $m_\phi < m_N$ or $m_N<m_\phi$. As discussed in \cite{Farzan:2014gza} the case with real scalar DM $\phi$ and pseudo-Dirac \footnote{A pseudo-Dirac particle corresponds to two almost degenerate Majorana particles, which behaves as a Dirac particle at the tree-level \cite{Wolfenstein:1981kw}.} $N$ is the only one for which couplings $g\sim\mathcal{O}(0.1)$ can produce the observed relic density of DM. For the other cases $g\ll\mathcal{O}(0.1)$, and the dip would not be detectable.


The  coupling  (\ref{eq:coupling}) introduces new decay modes for charged mesons, for instance $K^+\rightarrow e^+ + (\text{missing energy})$. Strong constraints on the coupling constants have been obtained, $|g_e|^2,|g_\mu|^2\lesssim 10^{-5}$ \cite{Bakhti:2017jhm}. Since the least constrained coupling is $g_\tau<1$, we assume $g_e=g_\mu=0$ and $g_\tau\neq 0$, which implies that $g_i = U_{\tau i}g_\tau$.

An interaction between dark matter and neutrinos can impact the evolution of primordial matter fluctuations \cite{Boehm:2004th,Mangano:2006mp,Campo:2017nwh}. Using cosmological data from Planck and the Lyman-$\alpha$ forest, the DM-neutrino elastic scattering cross section has an upper bound of
\begin{equation}
\sigma_{\text{DM}-\nu} < 10^{-36}\,\left(\frac{m_\text{DM}}{\text{MeV}}\right)\,\text{cm}^2,
\end{equation}
if the cross section is constant and
\begin{equation}
\sigma_{\text{DM}-\nu} < 10^{-48}\,\left(\frac{m_\text{DM}}{\text{MeV}}\right)\,\left(\frac{T_\nu}{T_0}\right)^2\,\text{cm}^2,
\end{equation}
if it scales as the temperature squared, with $T_0=2.35\times10^{-4}$ eV the temperature of the Universe today \cite{Wilkinson:2014ksa}. For the interaction considered in this work and the parameters we are using, the elastic scattering cross section is much lower than these two limits \cite{Campo:2017nwh}.

Dark matter in thermal equilibrium with neutrinos can increase the number of relativistic degrees of freedom, $N_{\text{eff}}$, in the early Universe if the DM particles annihilate into neutrinos after its decoupling from electrons at $T_{\text{dec}}\sim 2.3$ MeV. A larger $N_{\text{eff}}$ increases the expansion rate of the Universe, affecting the production of light elements during big bang nucleosynthesis (BBN) and erasing fluctuations on small scales \cite{Wilkinson:2016gsy}.


The relatively strong interactions between dark matter and neutrinos would maintain them in thermal equilibrium with each other out to radii where the temperature is low enough for the DM abundance to be suppressed. Hence dark matter particles do not carry away significant quantities of energy \cite{Bertoni:2014mva}. Because here dark matter does not couple to nucleons we do not expect a significant impact on supernovae cooling and the constraint in \cite{Fayet:2006sa} is avoided.

\section{Distortion of supernova neutrinos energy spectrum}
\label{sec:distortion}

\begin{figure}[t]
\begin{center}
\includegraphics[width=0.5\textwidth]{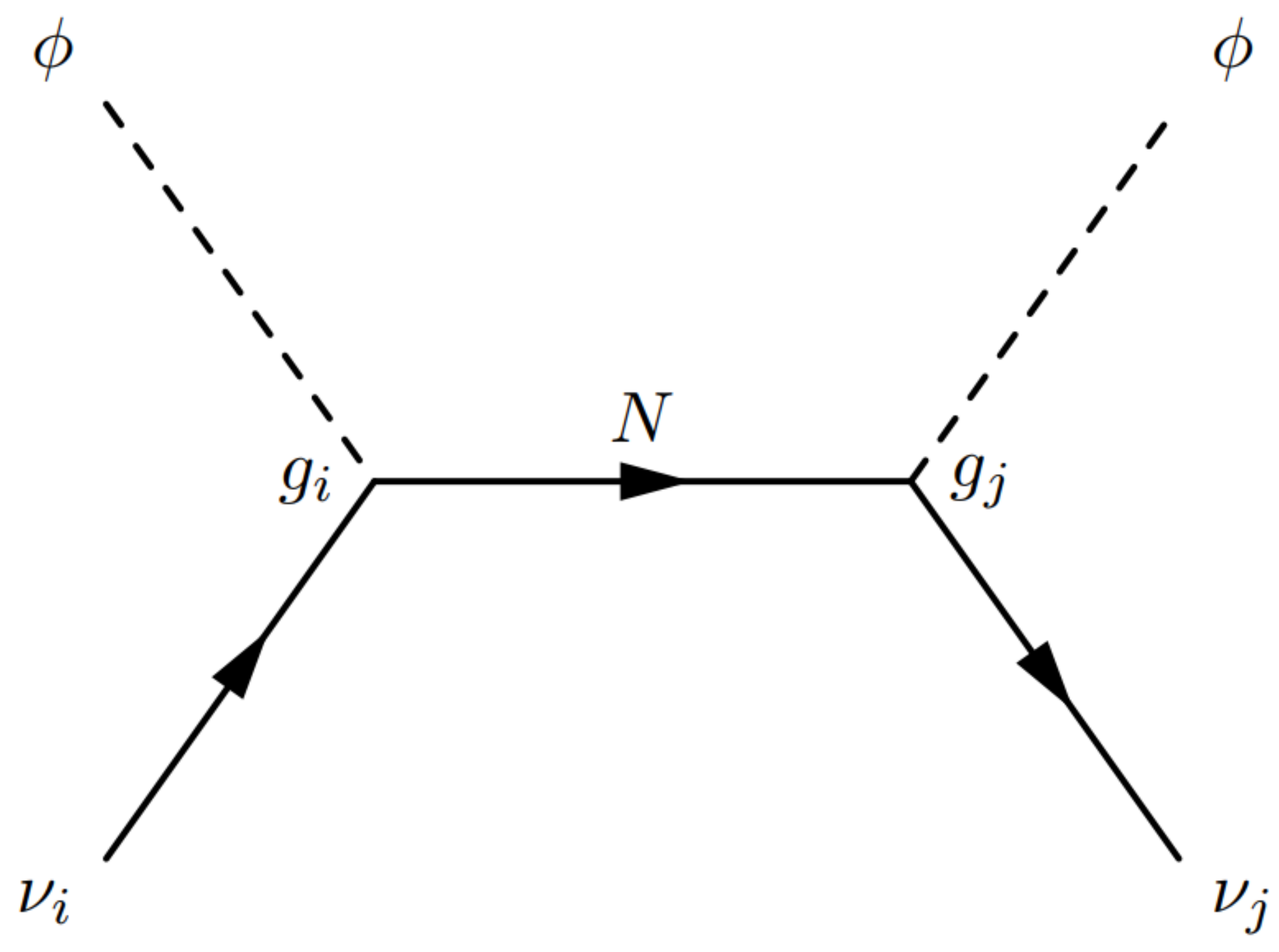}
\end{center}
\caption{Diagram for the s-channel of the $\nu_i\phi\rightarrow\nu_j\phi$ interaction.}
\label{fig:diagram}
\end{figure}

Neutrinos produced in a supernova interact with dark matter particles on their way to Earth. Here we neglect the non-resonant t-channel process and only consider the dominant s-channel in the total cross section, see figure \ref{fig:diagram}. When the center-of-mass energy of the DM-neutrino system is equal to the mass of the new fermion $N$ there is a resonant absorption, leading to a dip in the flux of neutrinos \cite{Farzan:2014gza}. The resonance energy in the laboratory frame is
\begin{equation}
E_r=\frac{m_N^2-m_\text{DM}^2}{2m_\text{DM}},
\end{equation}
where DM particles are considered to be at rest. An incoming neutrino with energy $E_\nu$ joins a DM particle to form an intermediate $N$ particle, which subsequently decays back into a DM particle plus a neutrino with energy $E_\nu^\prime$ given by
\begin{equation}
E_\nu^\prime = \frac{E_\nu}{2E_\nu+m_\text{DM}}\left[E_\nu(1+\text{cos}\theta)+m_\text{DM}\right],
\label{eq:energies}
\end{equation}
where $\theta$ is the emission angle of the neutrino with respect to its the incoming direction. 

Close to the resonance energy, the differential cross section of $\nu_i\phi \rightarrow N \rightarrow \nu_j\phi$ is given by
\begin{equation}
\frac{d\sigma_{ij}}{d\text{cos}\theta}=\frac{g_i^2g_j^2}{32\pi}\frac{(m_N^2-m_\text{DM}^2)^2}{m_N^2+m_\text{DM}^2}\frac{1+\text{cos}\theta}{(s-m_N^2)^2+\Gamma_N^2m_N^2},
\label{eq:diff_sigma}
\end{equation}
where $\Gamma_N$ is the decay width of the fermion $N$ and $s=2m_\text{DM}E_\nu+m_\text{DM}^2$ is the square of the center-of-mass energy. It follows that the total cross section is
\begin{equation}
\sigma_{ij}=\frac{g_i^2g_j^2}{16\pi}\frac{(m_N^2-m_\text{DM}^2)^2}{m_N^2+m_\text{DM}^2}\frac{1}{(s-m_N^2)^2+\Gamma_N^2m_N^2}.
\end{equation}


If $N$ decays predominantly into $\phi\nu$, its decay width is
\begin{equation}
\Gamma_N=\sum_i\frac{g_i^2}{16\pi}\frac{(m_N^2-m_\text{DM}^2)^2}{m_N^3}.
\end{equation}

As the neutrinos propagate through the DM halo, the processes $\nu_i\phi \rightarrow N \rightarrow \nu_j\phi$ may result in regeneration ($i=j$) and flipping ($i\neq j$) of mass eigenstates. We can describe the evolution of the flux as
\begin{equation}
\begin{split}
\frac{\partial F_{\nu_i}(x,E_\nu)}{\partial x}= & - n_\text{DM}(x) F_{\nu_i}(x,E_\nu) \sum_{j=1,2,3}\sigma_{ij}(E_\nu) \\&+ n_\text{DM}(x) \sum_{j=1,2,3} \int_{E_\nu}^\infty dE_\nu^\prime \frac{d\sigma_{ji}(E_\nu^\prime,E_\nu)}{dE_\nu}F_{\nu_j}(x,E_\nu^\prime).
\end{split}
\label{eq:propagation}
\end{equation}
The first term on the right-hand side accounts for the effects of absorption ($\nu_i\phi\rightarrow N$), while the second term stands for the regeneration and flipping ($N \rightarrow \nu_j\phi$). 



In principle we should solve a system of three coupled versions of equation (\ref{eq:propagation}), one for each mass eigenstate $\nu_i$. However, the mean free path of a SN neutrino in the Galactic halo is similar to the SN-Earth distance, $\lambda_{\nu}\sim d_\text{SN}$, and the suppression of the flux occurs in the same region of the spectra throughout the propagation because we are not considering evolution of the neutrinos through high redshifts. We then approximate the solution by first calculating the suppression of the flux as
\begin{equation}
F_{\nu_i}^\text{suppressed} = F_{\nu_i}^\text{SN}\,\text{exp}\left[-\sigma_{ij}\int_0^{d_\text{SN}}dx\,n_\text{DM}\right].
\label{eq:suppression}
\end{equation}
After that we redistribute the flux of absorbed neutrinos, $F_{\nu_i}^\text{SN}-F_{\nu_i}^\text{suppressed}$, with energies in the range given in (\ref{eq:energies}) with a probability proportional to $1+\text{cos}\theta$, see equation (\ref{eq:diff_sigma}) .


The flux of neutrinos of a given flavour from a typical SN is parameterised as
\begin{equation}
F_\alpha^\text{SN}(E_\nu)=\frac{L_{\nu_\alpha}}{\bar{E}_{\nu_\alpha}^2}\frac{(1+\beta_{\nu_\alpha})^{1+\beta_{\nu_\alpha}}}{\Gamma(1+\beta_{\nu_\alpha})}\left(\frac{E_\nu}{\bar{E}_{\nu_\alpha}}\right)^{\beta_{\nu_\alpha}}\text{exp}\left[-(1+\beta_{\nu_\alpha})\frac{E_\nu}{\bar{E}_{\nu_\alpha}}\right],
\label{eq:fluxSN}
\end{equation}
where $\bar{E}$ is the average energy and $\beta$ a numerical parameter. We assume $\bar{E}_{\nu_e}=10$ MeV, $\bar{E}_{\bar{\nu}_e}=12$ MeV, $\bar{E}_{\nu_x}=15$ MeV; $\beta_{\nu_e}=3$, $\beta_{\bar{\nu}_e}=3$, $\beta_{\nu_x}=2.4$; $L_{\nu_e}=L_{\bar{\nu}_e}=L_{\nu_x}=5\times 10^{52}$ ergs, where $x = \mu, \tau$ for neutrinos and antineutrinos. We use the Einsasto profile to describe the density of dark matter in the Galactic halo: 
\begin{equation}
\rho_\text{DM} (r) = 7.2\times 10^{-2}\,\text{GeV}\,\text{cm}^{-3}\,\text{exp}\left\{-\frac{2}{\alpha}\left[\left(\frac{r}{R_0}\right)^\alpha-1\right]\right\}
\label{eq:einasto}
\end{equation}
with $\alpha=0.15$ and $R_0=20$ kpc \cite{Einasto}. For our results, we assume the supernova occurs at a distance of 10 kpc and, in galactic coordinates, at longitude $l=45\degree$ and latitude $b=0\degree$, so in the disk but not particularly close to the centre of the Milky Way. Neutrinos from supernovae which occur close to or on the other side of the Galactic centre would experience much more absorption.
\begin{table}[t]
\centering
\label{my-label}
\begin{tabular}{c|c|c|c|c|c|c|c|c}
 $\bar{E}_{\nu_e}$ & $\bar{E}_{\bar{\nu}_e}$ & $\bar{E}_{\nu_x}$ & $\beta_{\nu_e}$ & $\beta_{\bar{\nu}_e}$ & $\beta_{\nu_x}$ & $d_\text{SN}$ & Latitude & Longitude \\ \hline
 10 MeV & 12 MeV & 15 MeV & 3 & 3 & 2.4 & 10 kpc & 45\degree & 0\degree \\
\end{tabular}
\caption{Parameters for the supernova neutrino energy fluxes considered in our calculations, see equation (\ref{eq:fluxSN}). The luminosities for all flavours are set as $L_{\nu_e}=L_{\bar{\nu}_e}=L_{\nu_x}=5\times 10^{52}$ ergs. The distance to the supernova is $d_\text{SN}$ and the angular position is given in galactic coordinates.}
\label{tab:parameters}
\end{table}

An important constraint comes from the observation of neutrinos from SN1987A. Since there is a good agreement between the expected and observed neutrino flux and energy spectrum, we assume
that those neutrinos were not significantly absorbed by dark matter. For neutrinos emitted from the Large Magellanic Cloud, equations (\ref{eq:suppression}) and (\ref{eq:einasto}) give the upper bound
\begin{equation}
\sigma_{ij}\lesssim 6.85\times 10^{-26}\,\text{cm}^2\,\left(\frac{m_\text{DM}}{\text{MeV}}\right).
\end{equation}
For our parameters, this condition is satisfied except near resonance, producing a dip in the neutrino spectrum. 


In the production region within the supernova, flavour eigenstates coincide with the mass eigenstates. For a slowly varying density and matter potential, the transitions between mass eigenstates are suppressed and the fluxes of mass eigenstates are directly related to the flavour fluxes at production:
\begin{equation}
\begin{split}
&F_{\bar{\nu}_1}^\text{SN}=F_{\bar{\nu}_e}^0, F_{\bar{\nu}_2}^\text{SN}=F_{\bar{\nu}_3}^\text{SN}=F_{\bar{\nu}_x}^0\,\,\,(\text{normal ordering});\\
&F_{\bar{\nu}_1}^\text{SN}=F_{\bar{\nu}_2}^\text{SN}=F_{\bar{\nu}_x}^0,F_{\bar{\nu}_3}^\text{SN}=F_{\bar{\nu}_e}^0\,\,\,(\text{inverted ordering}).
\end{split}
\label{eq:fluxes_SN}
\end{equation}

\begin{figure*}[t]
\includegraphics[width=0.48\textwidth]{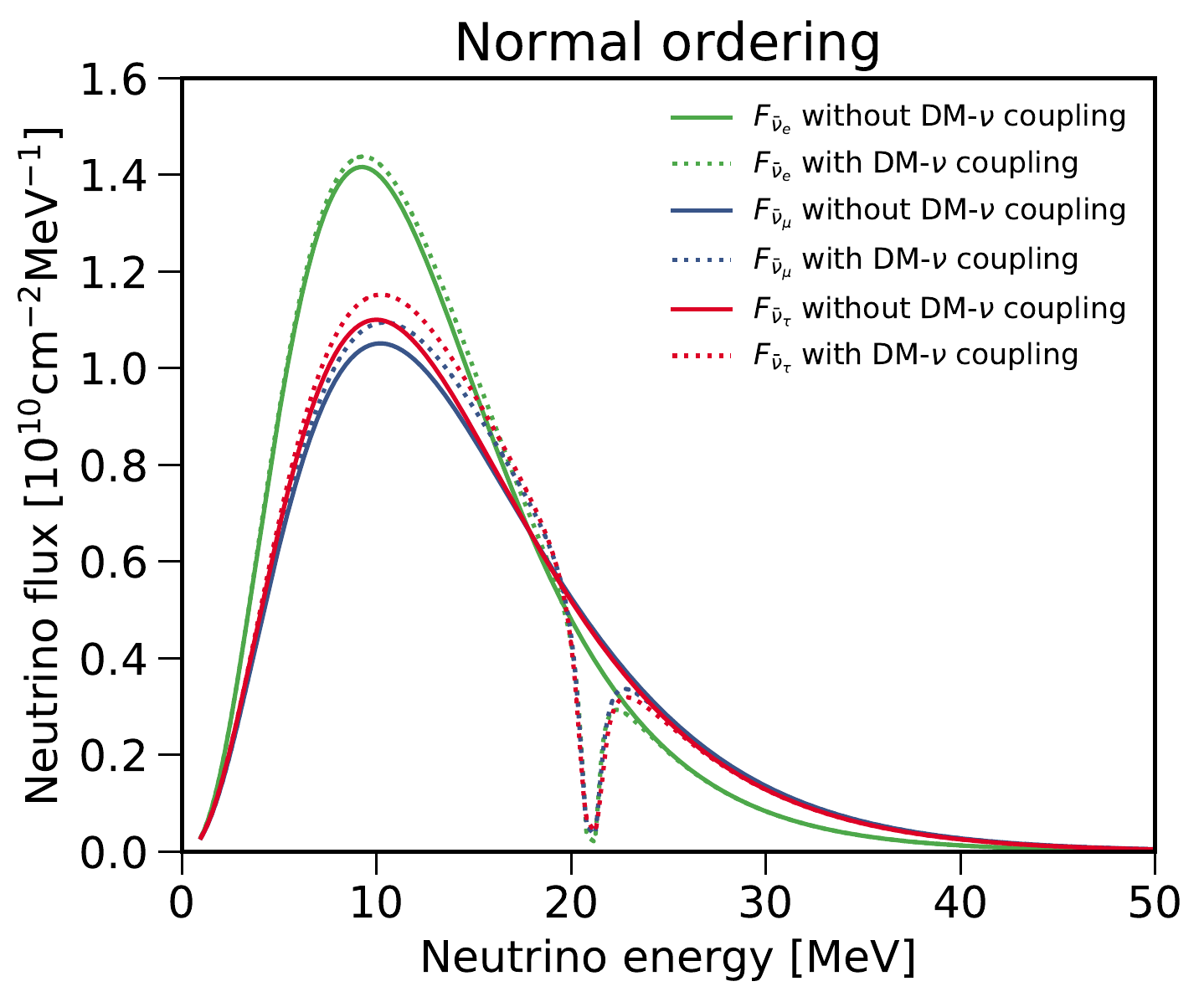}
\includegraphics[width=0.48\textwidth]{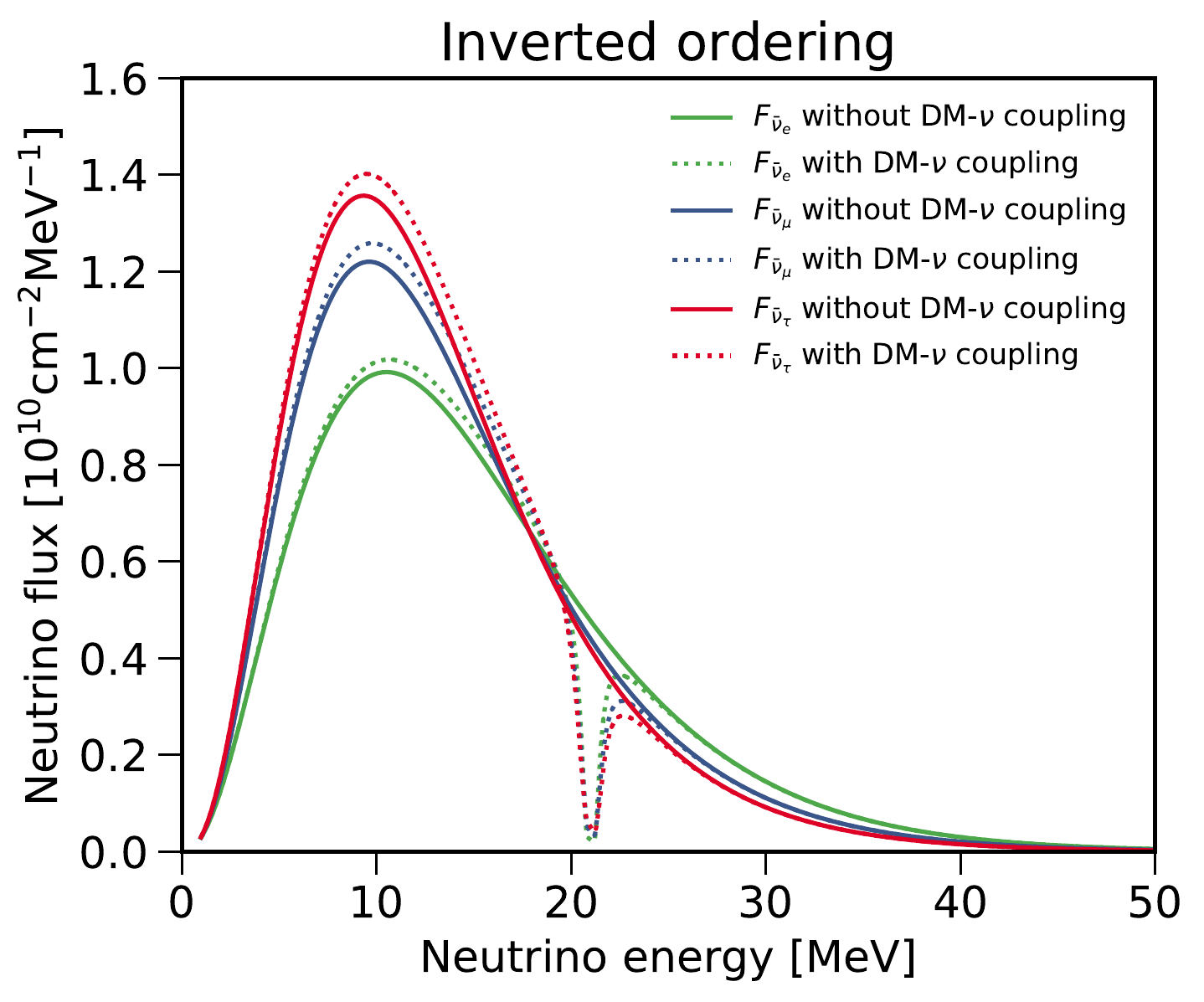}
\caption{Flux of antineutrinos on Earth from a typical supernova at a distance of 10 kpc. We assume $m_\phi=8$ MeV, $m_N=20$ MeV and $g_\tau=0.5$, for normal ordering (left panel) and inverted ordering (right panel).}
\label{fig:fluxes}
\end{figure*}

Supernovae neutrinos propagate without coherence to Earth so oscillations in vacuum are irrelevant \cite{Franarin:2017jnd}. The fluxes of neutrinos in the flavour basis at Earth, apart from the geometrical factor $(4\pi d_\text{SN}^2)^{-1}$ , are then given by
\begin{equation}
F_{\nu_\alpha}=\sum_{i=1}^4|U_{\alpha i}|^2F_{\nu_i},
\label{eq:flux_earth}
\end{equation}
with $F_{\nu_i}$ the mass eigenstates fluxes and $U_{\alpha i}$ the PMNS neutrino mixing matrix.

Let us take the electron antineutrinos as an example since the inverse beta decay is the main detection channel on liquid scintillators for the energies typical of supernovae neutrinos. From equation (\ref{eq:flux_earth}) we have
\begin{equation}
\begin{split}
&F_{\bar{\nu}_e}=0.67\,F_{\bar{\nu}_e}^0+0.33\,F_{\bar{\nu}_x}^0\,\,\,(\text{normal ordering});\\
&F_{\bar{\nu}_e}=0.02\,F_{\bar{\nu}_e}^0+0.98\,F_{\bar{\nu}_x}^0\,\,\,(\text{inverted ordering}).
\end{split}
\end{equation}
For the inverted ordering case, the flux of electron antineutrinos will have a hotter spectrum, corresponding to that of the original $\bar{\nu}_x$ flavour. 

Figure \ref{fig:fluxes} shows the fluxes of antineutrinos on Earth from a SN with the parameters in Table \ref{tab:parameters}. We assume $m_\phi=8$ MeV, $m_N=20$ MeV and $g_\tau=0.5$. Note that despite taking $g_e=g_\mu=0$, the electron and muon antineutrinos spectra are distorted due to neutrino mixing. We note that the dip has a width of $\mathcal{O}(1)$ MeV, then it is essential to use a detector with excellent energy resolution. JUNO is a future neutrino detector that can accomplish this.

\section{Detection with the JUNO Experiment}
\label{sec:detection}

\begin{figure}[t]
\begin{center}
\includegraphics[width=0.6\textwidth]{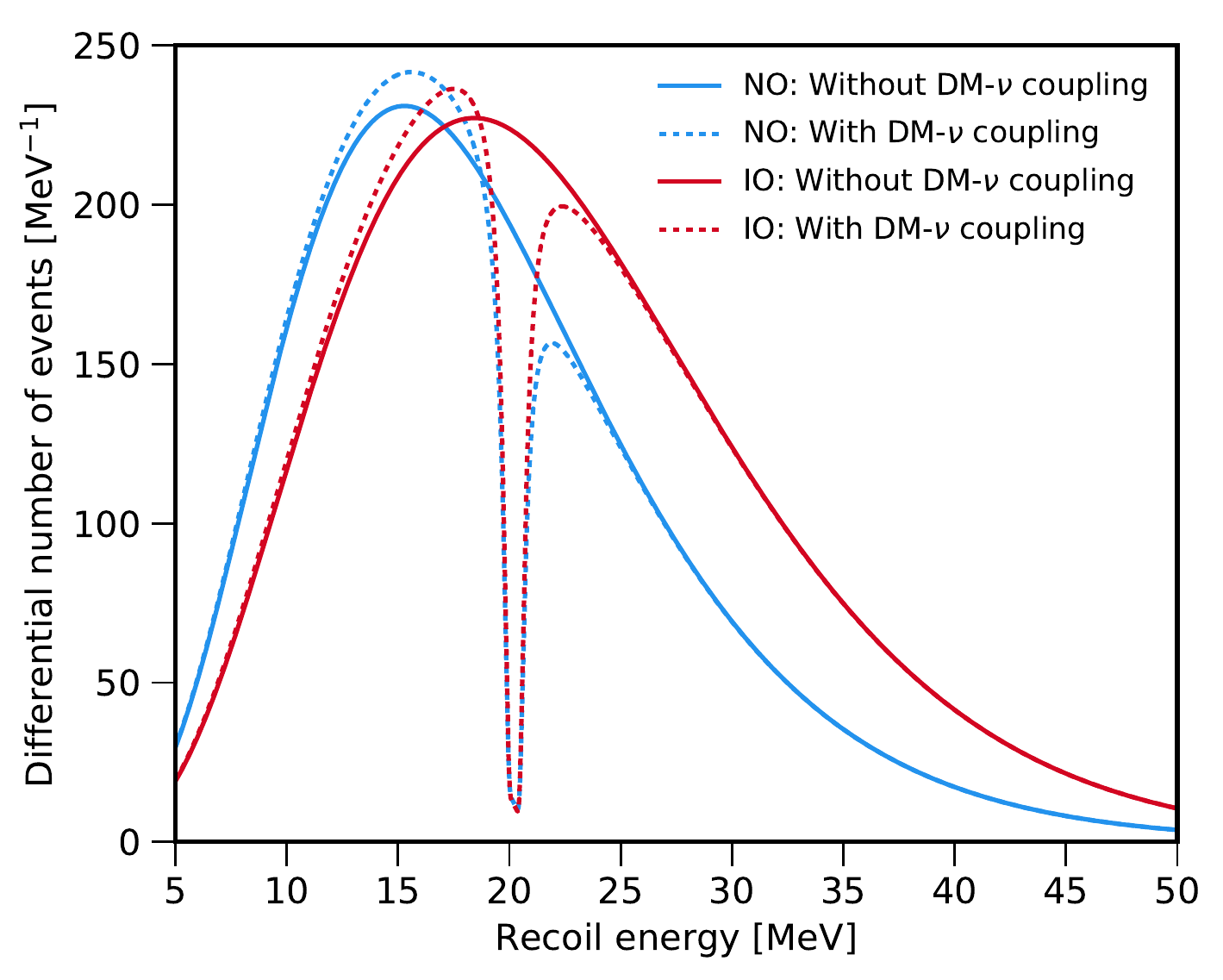}
\caption{Differential number of electron antineutrinos observed in JUNO for a typical supernova at a distance of 10 kpc, as a function of the detected positron recoil energy. We assume $m_\phi=8$ MeV, $m_N=20$ MeV and $g_\tau=0.5$, for normal (NO) and inverted (IO) orderings.\label{fig:juno}}
\end{center}
\end{figure}


JUNO is a 20 kton liquid scintillator detector being built in China which is expected to start data taking in 2020. Its main goal is to determine the neutrino mass ordering from the oscillation pattern of electron antineutrinos generated by two nearby nuclear power plants. The planned energy resolution of $3\%/\sqrt{E\,[\text{MeV}]}$ is one of the features that will allow the experiment to do so with a statistical significance of 3-4 $\sigma$ within six years of running \cite{An:2015jdp,Salamanna:2018pal}.

The inverse beta decay channel is the most important one for the detection of supernova neutrinos in liquid scintillator detectors. In this reaction,
\begin{equation}
\bar{\nu}_e+p\rightarrow n+e^+,
\end{equation}
the neutrino energy threshold is $E_\nu^{\text{th}}=m_n-m_p+m_e\approx 1.8$ MeV. The positron annihilates with an ambient electron into photons with energy $E_\gamma = E_{e^+} +m_e$ giving rise to a prompt signal. The energy of the interacting antineutrino can be reconstruted from the energy of the photons directly via $E_\nu = E_{e^+} + (m_n-m_p) \approx E_\gamma + 0.8$ MeV, neglecting the kinetic energy of the outgoing neutron. The neutron is captured by a proton $\sim 200 \mu s$ later and releases a 2.2 MeV photon. The coincidence of the prompt-delayed signal pair greatly reduces backgrounds. 

The total cross section for inverse beta decay is
\begin{equation}
\sigma(E_\nu)=\left[9.52\times 10^{-44}\,\text{cm}^2(E_\nu-1.3\,\text{MeV})^2\right]\left(1-7\frac{E_\nu}{m_p}\right)
\end{equation}
where $m_p$ is the proton mass \cite{Beacom:2010kk}. This approximation is very accurate at low energies ($E_\nu\leq 60$ MeV) thus can be safely used for supernova neutrino analyses \cite{Strumia:2003zx}. Figure \ref{fig:juno} shows the differential number of electron antineutrinos observed in JUNO from a SN with the parameters in Table \ref{tab:parameters} for normal and inverted orderings assuming $m_\phi=8$ MeV, $m_N=20$ MeV and $g_\tau=0.5$.

\section{Results}
\label{sec:results}

\begin{figure*}[t]
\includegraphics[width=0.49\textwidth]{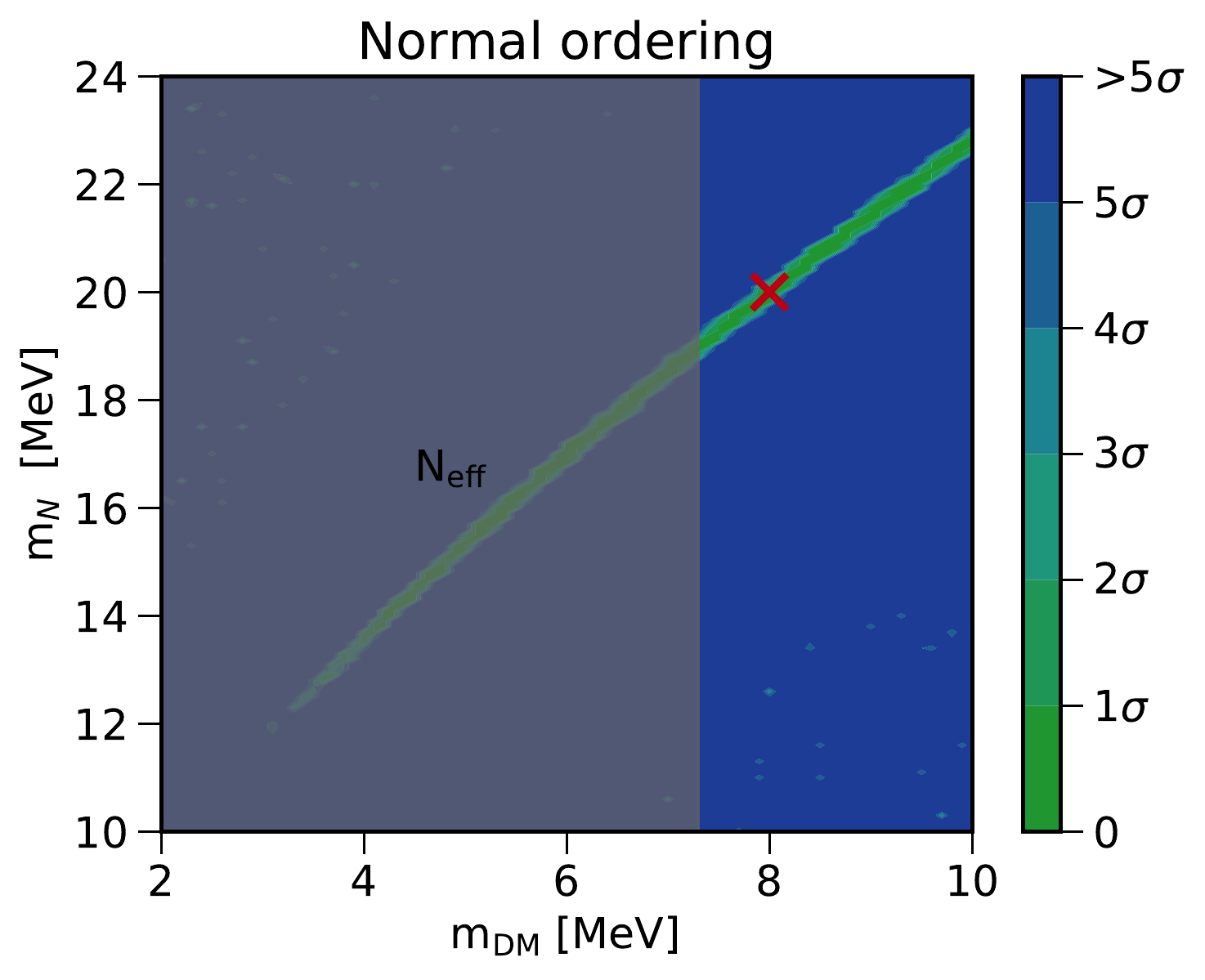}
\includegraphics[width=0.49\textwidth]{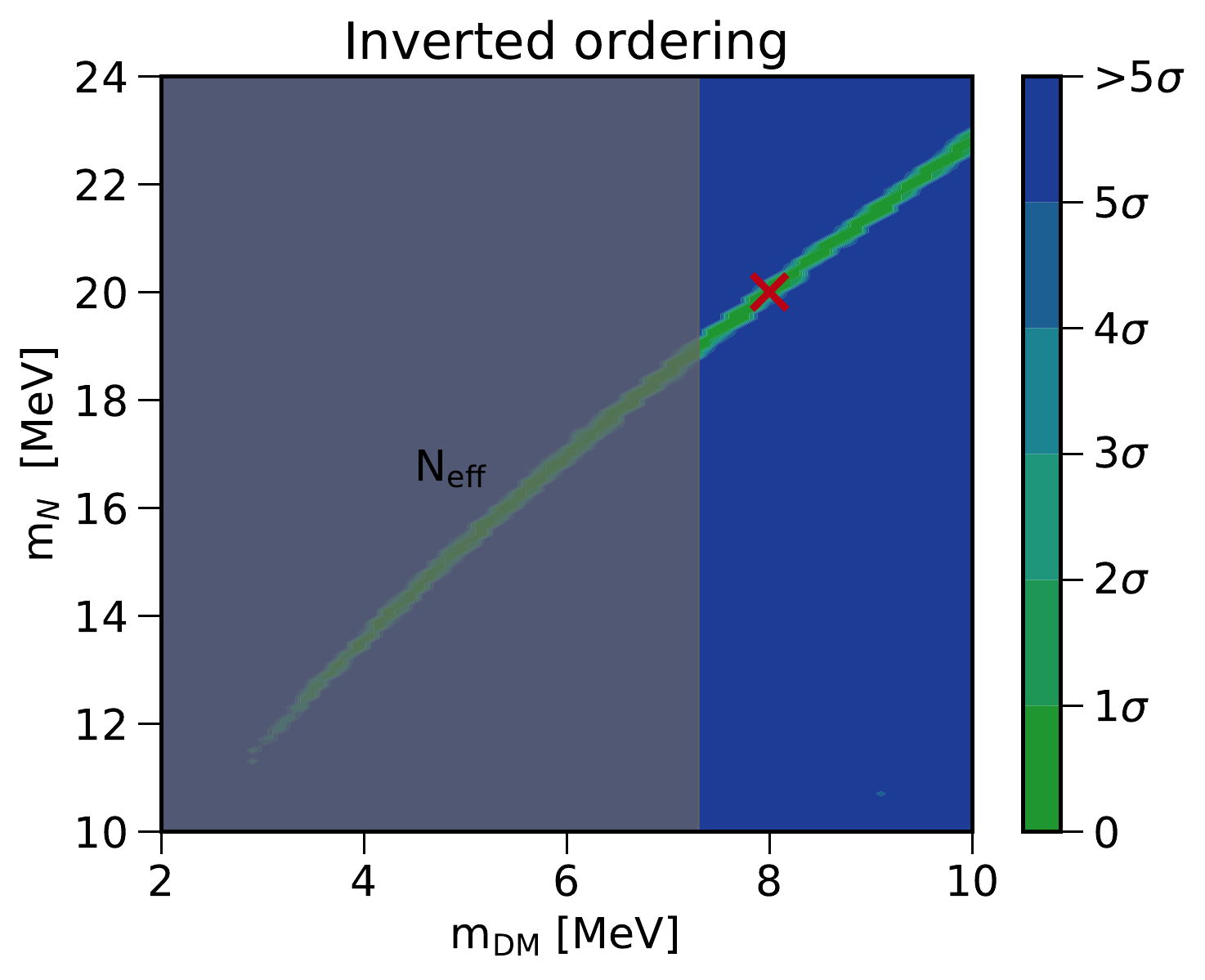}
\caption{Deviation from best fit for different dark matter and mediator masses for normal ordering (left panel) and inverted ordering (right panel). The fiducial values $m_\phi=8$ MeV, $m_N=20$ MeV and $g_\tau=0.5$ are marked in red and the grey band represents cosmological bounds assuming DM and neutrinos were in equilibrium in the early Universe \cite{Wilkinson:2016gsy}.\label{fig:stats}}
\end{figure*}

In order to find out if JUNO is sensitive to this interaction, we performed the following analysis.  We chose fiducial values for the coupling and the masses of the dark matter and the mediator such that the resonance occurs in the interesting region for supernova neutrinos. In particular we took $g_\tau=0.5$, $m_\phi=8$ MeV and $m_N=20$ MeV. We then generated events making these assumptions and saw how well we can recover the values that we put into the event generator using the expected detector response for JUNO. 

We consider inverse beta decay events only and group them in 0.2 MeV bins. The total likelihood of a given set of observed data is given by
\begin{equation}
\mathcal{L}= \prod_i \text{e}^{-\lambda_i}\frac{\lambda_i^{k_i}}{k_i!},
\end{equation}
where $\lambda_i$ and $k_i$ are the expected observed number of events on the bin $i$, respectively.

The goodness of the fit is associated with the probability
\begin{equation}
p=\int_0^{-2\text{log}\left(\mathcal{L}/\mathcal{L}_\text{max}\right)}f_{\chi^2}(x;k)\,\text{d}x,
\end{equation}
where $f_{\chi^2}$ is the pdf of a chi-squared random variable with $k$ degrees of freedom, and $\mathcal{L}_\text{max}$ is the maximum likelihood. These probabilities can be translated into the number of standard deviations from the best fit.

Figure \ref{fig:stats} shows the goodness of fit for the observed spectrum of neutrinos from a Galactic SN in JUNO fixing $g_\tau=0.5$. The cross represents the fiducial values for the parameters, $m_\phi=8$ MeV and $m_N=20$ MeV. There is a degeneracy in the reconstruction of the dark matter mass and the mediator fermion mass for those mass combinations which lead to a dip at the same resonance energy with approximately the same width.

\begin{figure}[t]
\begin{center}
\includegraphics[width=0.6\textwidth]{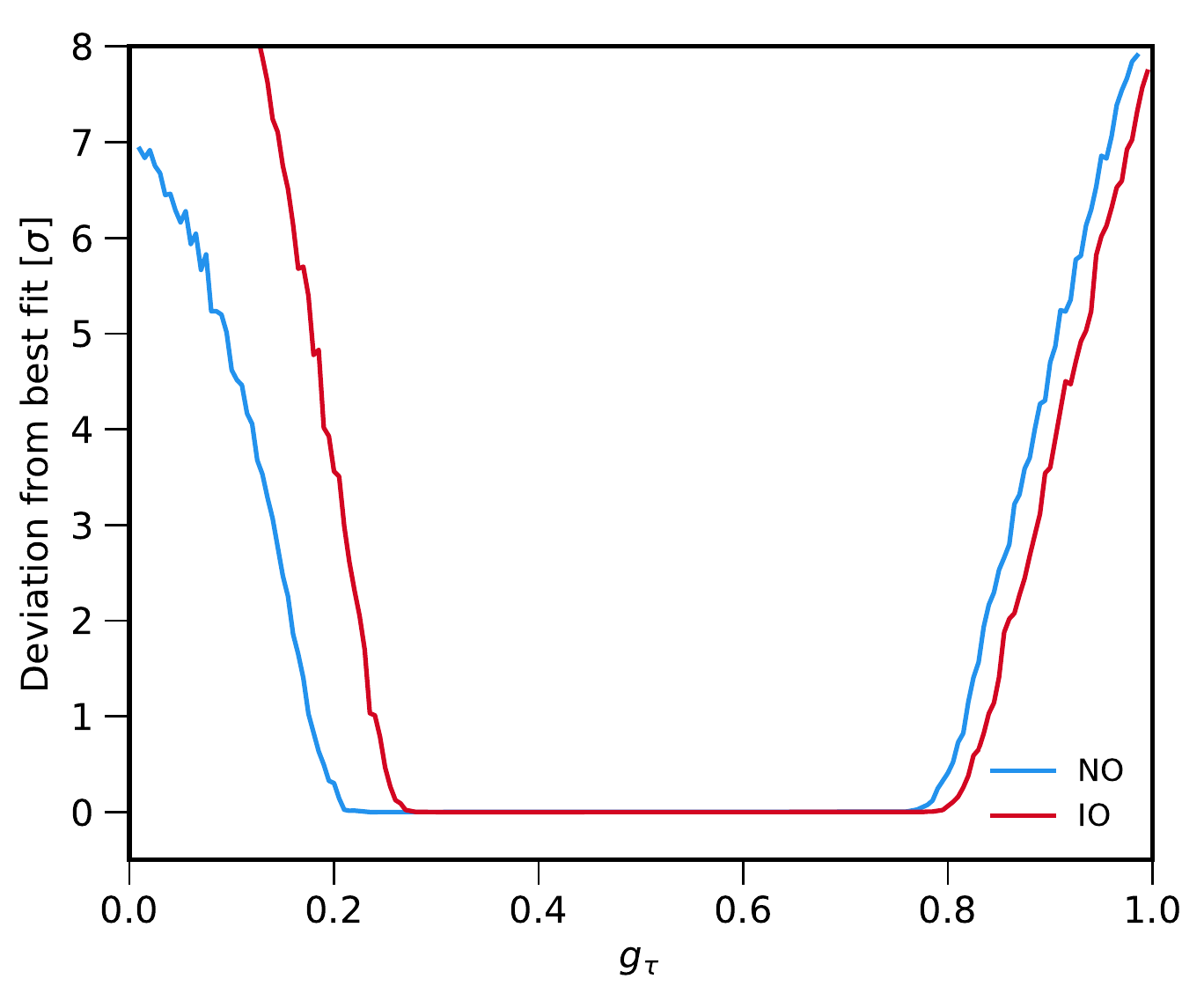}
\caption{Deviation from the best fit fixing $m_\phi=8$ MeV, $m_N=20$ MeV and varying $g_\tau$ for normal and inverted orderings. The slight differences in behaviour for the two cases are due to different spectral shapes in the two scenarios, as explained in the text.\label{fig:g} }
\end{center}

\end{figure}

In figure \ref{fig:g} we fix $m_\phi=8$ MeV and $m_N=20$ MeV and vary $g_\tau$. There is a 50\% uncertainty in the determination of the coupling constant. As can be seen in figure \ref{fig:juno}, the resonance energy for these parameters occurs close to the peak of the spectrum in the inverted ordering case, while for the normal ordering situation, the dip occurs where there are fewer neutrinos, explaining the difference between the two curves.  The sensitivity to couplings is a result of the excellent energy resolution of the JUNO detector.  Improved energy resolution would result in sensitivity to even smaller couplings.

\begin{figure}[t]
\begin{center}
\includegraphics[width=0.6\textwidth]{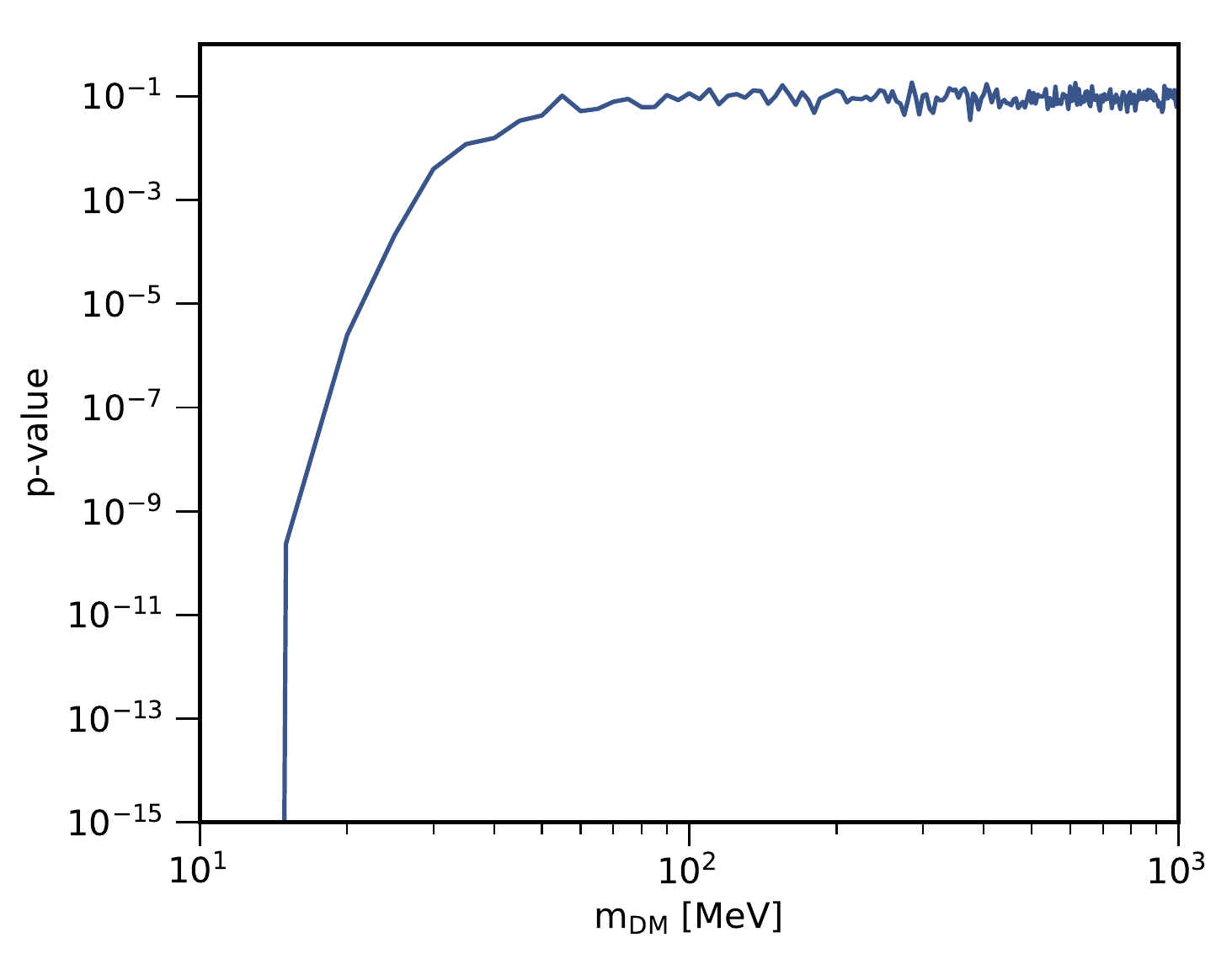}
\caption{Deviation from the expected flux without DM-neutrino coupling for different dark matter masses assuming $g_\tau$=0.5, $E_\text{res}$=20 MeV and normal ordering of neutrino masses.  This plot shows that for heavier dark matter, this detection strategy will start to become inefficient.\label{fig:sensitivity}}
\end{center}

\end{figure}

In figure \ref{fig:sensitivity} we fix $g_\tau$=0.5 and $E_\text{res}$=20 MeV for normal ordering and calculate the deviation from the expected flux without DM-neutrino coupling for different dark matter masses. We can see that it is possible to detect the effects of this coupling for DM masses up to around 20 MeV.  This lack of sensitivity to high masses is because even though it is always possible to arrange a small mass difference between the dark matter mass and the mediator such that the resonance is in the region of interest, once the mass of the mediator becomes very large, the width of the resonance becomes smaller than the energy resolution of JUNO, meaning the dip become undetectable.

We can compare our results with those of reference \cite{Farzan:2014gza} where the authors look for absorption features in the diffuse supernova neutrino background.  The first thing to say is that the cosmological constraints presented in \cite{Wilkinson:2016gsy} actually rule out the fiducial parameters chosen in that paper, however we would still expect their method to be sensitive for different masses which are not yet ruled out.  They require Hyper-Kamiokande to be built which probably will not take place until at least 2025 and then require a period of data taking which depends strongly upon the final target mass of the experiment. We need JUNO for our analysis which is currently under construction and is due to start taking data possibly as early as 2020. However we also require a galactic supernova, the time schedule for which is a lot more uncertain, although most estimates suggest that we are overdue a nearby explosion.

\section{Conclusions}
\label{sec:conclusions}

Very little is known about the nature of dark matter other than the fact that its coupling to quarks, charged leptons and the bosons of the Standard Model must be very small. The same cannot be said of the couplings between dark matter and the neutrinos where many different interactions are still allowed. In this work we have considered the possibility of detecting DM-neutrino interactions through their effect upon the spectrum of neutrinos observed during a Galactic supernova. For a simple model where neutrinos couple to scalar dark matter through a new fermionic mediator we have shown that it would be viable to confirm such an interaction with future neutrino detectors.

In particular we have focused on the JUNO detector due to its extremely good energy resolution that enables one to search for the s-channel resonance (figure \ref{fig:juno}). There is a degeneracy in the reconstruction of the dark matter mass and the mediator fermion mass because they would lead to a dip at the same resonance energy with about the same width (figure \ref{fig:stats}). The effect is detectable in JUNO for dark matter masses up to around 20 MeV beyond which the decay width of the dark mediator would become too small for the energy resolution of the JUNO detector, which is envisaged to have the best energy resolution of any upcoming detector.

It would be interesting to look at other models of dark matter-neutrino interactions that might give rise to a similar effect, however, as alluded to in the text and explained in more detail in \cite{Farzan:2014gza}, it is only in the class of models mentioned here with a pseudo-Dirac mediator and real scalar dark matter that such a large coupling can be compatible with relic abundance.  Since a large coupling is required to obtain a detectable dip, other models which could lead to such an effect would be limited. Nevertheless it would be interesting to investigate in more detail a wider set of particle models and the constraints that could be obtained when a supernova goes off in the Milky Way.

\section*{Acknowledgements}
TF thanks support from CNPq SwB grant. The research leading to these results has been funded by the European Research Council through the project DARKHORIZONS under the European Union's Horizon 2020 program (ERC Grant Agreement no.648680).  The work of MF was also supported partly by the  STFC Grant  ST/L000326/1.

\bibliographystyle{JHEP}

\providecommand{\href}[2]{#2}\begingroup\raggedright\endgroup

\end{document}